\documentclass[aps,pre,twocolumn,showpacs,a4paper,superscriptaddress]{revtex4-1}
\usepackage{graphicx,bm,amsmath,dcolumn,amssymb}
\usepackage{pstricks,pst-node,pst-text,pst-3d,pst-plot}
\graphicspath{{figures}}
\usepackage{geometry,hyperref}
\begin{document}
\newcommand{\half}{\frac{1}{2}}
\newcommand{\ith}{^{(i)}}
\newcommand{\im}{^{(i-1)}}
\newcommand{\gae}
{\,\hbox{\lower0.5ex\hbox{$\sim$}\llap{\raise0.5ex\hbox{$>$}}}\,}
\newcommand{\lae}
{\,\hbox{\lower0.5ex\hbox{$\sim$}\llap{\raise0.5ex\hbox{$<$}}}\,}
\newcommand{\be}{\begin{equation}}
\newcommand{\ee}{\end{equation}}
\newcommand{\bea}{\begin{eqnarray}}
\newcommand{\eea}{\end{eqnarray}}

\title{Critical properties of the Hintermann-Merlini model}
\author{Chengxiang Ding}
\email{dingcx@ahut.edu.cn}
\affiliation{Department of Applied Physics, Anhui University of Technology, Maanshan 243002, China}
\author{Yancheng Wang}
\affiliation{Beijing National Laboratory for Condensed Matter Physics, Institute of Physics,
Chinese Academy of Sciences, Beijing 100190, China}
\author{Wanzhou Zhang} 
\affiliation{College of Physics and Optoelectronics, Taiyuan University of Technology, Shanxi 030024, China}
\author{ Wenan Guo}
\email{waguo@bnu.edu.cn}
\affiliation{Physics Department, Beijing Normal University, Beijing 100875, China}

\date{\today} 

\begin{abstract}

Many critical properties of the Hintermann-Merlini model 
are known exactly through the mapping to the eight-vertex model. 
Wu [J. Phys. C {\bf 8}, 2262 (1975)] calculated the spontaneous magnetizations of the model on two 
sublattices by relating them to the conjectured spontaneous magnetization and
polarization of the eight-vertex model, respectively. The latter conjecture remains unproved.
In this paper, we numerically study the critical properties of the model by means of a finite-size scaling analysis 
based on transfer matrix calculations and Monte Carlo simulations. 
All analytic predictions for the model are confirmed by our numerical results.
The central charge $c=1$ is found for critical manifold investigated. 
In addition, some unpredicted geometry properties of the model are studied. 
Fractal dimensions of the largest Ising clusters on two sublattices are determined. The fractal dimension of the largest Ising cluster on the sublattice A 
takes fixed value $D_{\rm a}=1.888(2)$, while that 
for sublattice B varies continuously with the parameters of the model.
 \end{abstract}
\maketitle 
\section{Introduction}

The exact solutions\cite{Ising,YangIsing} of the two-dimensional (2D) Ising 
model significantly promote the research of phase transitions and critical phenomena. 
After that, the Ising model becomes one of the most famous lattice models in statistical physics.
Another famous Ising system, known as the Baxter-Wu model\cite{bw}, is defined on the triangular lattice with pure three-spin interactions.
The model was firstly proposed by Wood and Griffiths\cite{wood} and exactly solved by Baxter and Wu\cite{bw}
by relating the model with the coloring problem on the honeycomb lattice.  
The solution gives the 
critical exponents $y_t=3/2$ ($\alpha=2/3$) and $y_h=15/8$ ($\eta=1/4$), which are exactly the same as those of the 4-state Potts 
model\cite{potts,wfypotts}, which can be derived via the Coulomb gas theory\cite{denNijs1,denNijs2,cg}.
 This means that the Baxter-Wu model belongs to the
universality class of the 4-state Potts model. 
The critical properties of the 4-state Potts model are modified by logarithmic corrections\cite{log2,log}
due to the second temperature field, which is marginally irrelevant\cite{NBRS,log2}.
With the two leading temperature fields simultaneously vanishing, the leading critical singularities of the Baxter-Wu model\cite{dengbw} 
do not have logarithmic factors. 
Deng {\it et al.} 
generalize the Baxter-Wu model in \cite{dengbw} where the spins are allowed to be $q$ states ($q$ can be larger than 2) and the up- and 
down-triangles can have different coupling constants. Both generalizations lead to discontinuous phase transitions.

\begin{figure}[htpb]
\includegraphics[scale=1.0]{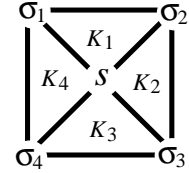}
\caption{(Color online) Definition of Hintermann-Merlini (HM) model: the lattice consists of two sublattices, the spin on sublattice A is denoted
$\sigma$ and the spin on sublattice B is denoted $s$; for a square unit cell, there are four spins $\sigma_1, \sigma_2,\sigma_3$, and $\sigma_4$ on  
sublattice A and one spin $s$ on sublattice B. }
\label{uj}
\end{figure}
\begin{figure}[htpb]
\includegraphics[scale=0.4]{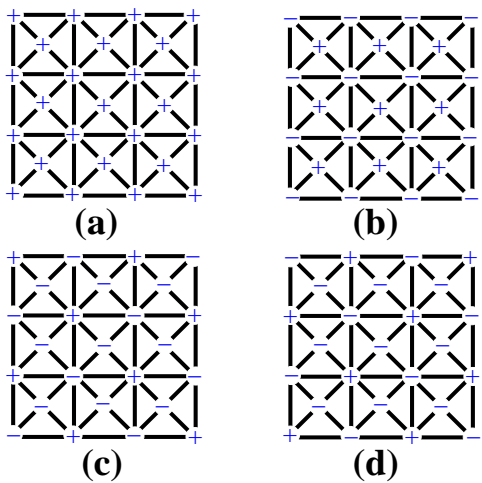}
\caption{(Color online) Four ground states of the ferromagnetic HM model.}
\label{ground}
\end{figure}

In 1972, Hintermann and Merlini\cite{hintermann} considered an Ising system on the 
Union-Jack lattice (as shown in Fig. \ref{uj})
\begin{eqnarray}
-\frac{H}{k_{\rm B} T}=\sum\limits_{\Box } s(K_1\sigma_1\sigma_2+K_2\sigma_2\sigma_3+K_3\sigma_3\sigma_4 \nonumber \\ 
+K_4\sigma_4\sigma_1),\label{wubw}
\end{eqnarray}
where the sum takes over all the square unit cells and $K_i$ ($i=1,2,3,4$) are the coupling constants.
The model is similar to the Baxter-Wu model in the sense that both are Ising 
systems with pure three-spin interactions. However 
the critical properties of the HM model are much more 
complicated. 
For the ferromagnetic case ($K_i>0$), the model has four-fold degenerate ground states, as shown in Fig. \ref{ground}. 
The nature of the phase transition breaking the $Z_4$ symmetry of the order 
parameter cannot be determined by the dimensionality of the system and the 
symmetry properties of the ground states\cite{Suzuki}. The exponents may vary with some tuning parameter without
changing the symmetry of the order parameter.
Such behavior has been found, e.g., in the Ashkin-Teller model (AT)
\cite{AT,ATMC}, the eight-vertex model \cite{8vertex,exactbook}, 
the 2D $XY$ model in a four-fold anisotropic field\cite{XYh}, and the ferromagnetic Ising model 
with antiferromagnetic next-nearest neighbor couplings on square lattice\cite{j1j2-henk,j1j2-sandvik,j1j2-guo}. 

Through mapping to the eight-vertex model\cite{8vertex,exactbook}, 
the free energy of the HM model has been found exactly\cite{hintermann}. 
The critical manifold and critical exponent $y_t$ varying with the ratio of 
couplings were obtained. 
Based on the conjectured spontaneous magnetization \cite{8v-mag} and
the spontaneous polarization \cite{8v-polar} of the eight-vertex model, Wu
calculated the spontaneous magnetizations of the model on two 
sublattices. The results show that the two magnetizations possess different 
critical exponents\cite{wubw}. 
The spontaneous magnetization of the eight-vertex
model was derived later \cite{8v-mag-proof}, however, the spontaneous polarization
remains a conjecture. 
It is thus very necessary to verify the results numerically.

In current paper, we study the critical behavior of the ferromagnetic HM 
model with $K_1=K_3>0$ and $K_2=K_4>0$ numerically. 
The numerical procedure  includes transfer matrix calculations and  
Monte Carlo simulations. The critical properties we 
studied includes not only the verification of the analytic predictions,
but also the fractal structure of spin clusters, which is studied for the 
first time.

The paper is organized in the following way: In Sec. \ref{sec_exact}, we summarize 
the theoretical results of the model.
In Sec. \ref{sec_TM}, we introduce the transfer matrix method 
and present the associated numerical results. 
In Sec. \ref{sec_MC}, we describe the algorithm used in Monte Carlo 
simulations and give the associated numerical results. 
In Sec. \ref{sec_FC}, we define Ising clusters on two sublattices and 
numerically determine the fractal dimensions of the corresponding largest 
clusters, respectively.
We summarize in Sec. \ref{sec_Con}. 

\section{Exact solutions}
\label{sec_exact}
 In this section, we summarize the analytical results for the HM model\cite{hintermann,wubw}.

\begin{figure}[htpb]
\includegraphics[scale=0.45]{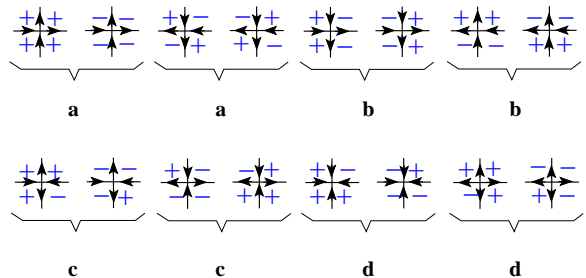}
\caption{(Color online) The mapping of the HM model to the eight-vertex model and the Boltzmann weights of 
the vertices.  The Ising spins shown here are the spins on sublattice A, namely, the $\sigma$ spins.}
\label{vertex}
\end{figure}

Assign arrows between the nearest-neighboring Ising spins of sublattice A (Fig. \ref{vertex}): if the two spins beside the 
arrow are the same, the arrow is rightward (or upward), otherwise it is leftward (or downward).
There is a two-to-one correspondence between the Ising spin configurations and the arrow configurations.

After taking sum over the spin $s$ (in the center of the unit cell), the 
Boltzmann weights of the vertices are
\begin{eqnarray}
a&=&\cosh(K_1+K_2+K_3+K_4),\nonumber\\
b&=&\cosh(K_1-K_2+K_3-K_4),\nonumber\\
c&=&\cosh(K_1-K_2-K_3+K_4),\nonumber\\
d&=&\cosh(K_1+K_2-K_3-K_4).
\end{eqnarray}
Thus the model is mapped to the symmetric eight-vertex model\cite{8vertex,exactbook}, which has been exactly solved by Baxter. 
For the ferromagnetic case considered in current paper, the critical manifold 
is given by 
\begin{eqnarray}
a=b+c+d, 
\label{cp}
\end{eqnarray} 
which leads to $K_{\rm c}=\log(1+\sqrt{2})/2$ for the uniform case ($K=K_1=K_2=K_3=K_4$).

The singularity of the free energy density is governed by
\begin{eqnarray}
f_{\rm sing} \propto |T-T_{\rm c}|^{\pi/u}
\end{eqnarray}
when $\pi/2u$ is not an integer (The case that $\pi/2u$ = an integer is not 
considered in current paper).
$T_{\rm c}$ is the critical temperature and  $0\le u \le \pi$ is given by \cite{wubw}
\begin{eqnarray}
\cos u&=&-\tanh[\frac{1}{2}\log\frac{ab}{cd}]\Big|_{T_{\rm c}}\quad {\rm if}\  a>b,c,d, \nonumber\\
      &&~~~~~~~~~~~~~~~~~~~~~~~~~~~~~{\rm or \ } b>a,c,d  \label{cosu}\\
      &=&\tanh[\frac{1}{2}\log\frac{ab}{cd}]\Big|_{T_{\rm c}}\quad \quad{\rm if} \  d>a,b,c, \nonumber\\
&&~~~~~~~~~~~~~~~~~~~~~~~~~~~~~{\rm or \ } c>a,b,d.
\end{eqnarray}
For the ferromagnetic case that we consider, $u$ is determined by (\ref{cosu}).
Since $f_{\rm sing} \propto \xi^2$ and $\xi \sim |T-T_{\rm c}|^{-1/y_t}$, the critical exponent $y_t$ is thus found 
\begin{eqnarray}
y_t=\frac{2u}{\pi}. \label{ytexact}
\end{eqnarray}
For the uniform case, (\ref{ytexact}) gives $y_t=4/3$.

A very interesting critical property of the HM model is that the spontaneous magnetizations $M_{\rm A}$ and
$M_{\rm B}$ of the sublattice A and B possess different critical exponents: 
\begin{eqnarray}
M_{\rm A} &\propto& (T_{\rm c}-T)^{\beta_{a}}, \nonumber \\ 
M_{\rm B} &\propto& (T_{\rm c}-T)^{\beta_{b}}. 
\end{eqnarray}
The critical exponents $\beta_{a}$ and $\beta_{b}$ were obtained by Wu\cite{wubw}
\begin{eqnarray}
{\beta_{a}}&=&\frac{\pi}{16u},\nonumber\\
{\beta_{b}}&=&\frac{\pi-u}{4u}. \label{betaab}
\end{eqnarray}
(It should be noted that the author reversely wrote the two critical exponents in Eq. (16) of \cite{wubw}.)
According to the scaling law, this leads to two magnetic exponents $y_{h1}$ and $y_{h2}$
\begin{eqnarray}
y_{h1}&=2-\beta_{a} y_t&=\frac{15}{8},\nonumber\\
y_{h2}&=2-\beta_{b} y_t&=\frac{3\pi+u}{2\pi}.\label{yh12}
\end{eqnarray}
Here $y_{h1}$ has fixed value $15/8$, while $y_{h2}$ varies continuously with the parameters of the model.

In following sections, we will numerically study the critical properties of the model. The 
critical points and critical exponents will be numerically verified. 
Especially,  Wu's result of $\beta_{b}$ ($y_{h2}$) is based on 
the unproved conjecture of the spontaneous polarization\cite{8v-polar} of the
eight-vertex model, thus numerical verification is very necessary.
Our numerical studies are focused on the subspace that $K_1=K_3=K$ and $K_2=K_4=K'$. By defining $r \equiv K'/K$,
critical manifold and exponents are expressed as functions of $r$. It is worthy to notice that
there is a symmetry for the transformation $r \to 1/r$. At a special point $r^*=3.3482581805$ and its
 dual $1/r^*=0.2986627512$, $u=3 \pi/4$, 
the analytic results give $y_t=1/2, y_{h1}=15/8=y_{h2}$. This is the 4-state Potts point.

\section{Transfer matrix calculations}
\label{sec_TM}
\begin{figure}[htbp]
\includegraphics[scale=0.65]{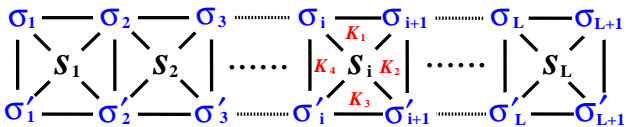}
\caption{(Color online) Definition of the row-to-row transfer matrix, where periodic boundary
 conditions are applied, namely $\sigma_{L+1}=\sigma_1$, $\sigma_{L+1}^\prime=\sigma_1^\prime$.}
\label{tm}
\end{figure}

As shown in Fig. \ref{tm}, we define the row-to-row transfer matrix 
\begin{eqnarray}
T_{\vec{\sigma},\vec{\sigma}^\prime}
&=&\sum\limits_{\{s\}}
\prod\limits_{i=1}^L 
\exp[s_i(K_1\sigma_i\sigma_{i+1}+K_2\sigma_{i+1}\sigma_{i+1}^\prime+ \nonumber \\
& & K_3\sigma_{i+1}^\prime\sigma_{i}^\prime+K_4\sigma_i^\prime\sigma_i )]\nonumber\\
&=&2^L\prod\limits_{i=1}^L
\cosh(K_1\sigma_i\sigma_{i+1}+K_2\sigma_{i+1}\sigma_{i+1}^\prime  \nonumber \\
&& +K_3\sigma_{i+1}^\prime\sigma_{i}^\prime+K_4\sigma_i^\prime\sigma_i ),
\end{eqnarray}
where $\vec{\sigma}=(\sigma_1,\sigma_2,\cdots,\sigma_L)$ and 
$\vec{\sigma}^\prime=(\sigma^\prime_1,\sigma^\prime_2,\cdots,\sigma^\prime_L)$ are the 
states of two neighboring rows, respectively. Here periodic boundary conditions are applied. 
For a system with $M$ rows, the partition sum is found to be 
\begin{eqnarray}
Z &=&{\rm Trace} (T^M), 
\end{eqnarray}
with the periodic boundary conditions $\vec{\sigma}_{M+1}=\vec{\sigma}_1$ applied. 
In the limit $M\rightarrow \infty$, the free energy density is determined by the leading eigenvalue $\Lambda_0$ of $T$
\begin{eqnarray}
f&=&\frac{1}{L}\log\Lambda_0 \label{free}.
\end{eqnarray}

For the HM model, the dimension of the matrix $T$ is $d_T=2^L$. 
To numerically calculate the eigenvalues of $T$, we used the sparse matrix technique,
which sharply reduces the requirement of computer memory for storing the matrix elements.
(For details of this technique, see \cite{henk1982,henk1989,henk1993,Qian2004}.)
We are able to calculate the eigenvalues of $T$ with $L$ up to 22. 
In our calculations, we restrict the system size $L$ to even values, because  
the ordered configurations, as shown in Fig. \ref{ground}, do not fit well in odd
systems.

The critical properties can be revealed by calculating three scaled gaps $X_i(K,L)$:
\begin{eqnarray}
X_i (K,L)=\frac{1}{2\pi}\log \big(\frac{\Lambda_0}{\Lambda_i}\big), \label{gap}
\end{eqnarray}
where $\Lambda_i (i=1,2,3)$ are three subleading eigenvalues of the matrix $T$, respectively.
According to the conformal invariance theory\cite{conformal}, the scaled gap $X_i(K,L)$ is related with a correlation 
length $\xi_i(K,L)$ 
\begin{eqnarray}
X_i(K,L)=\frac{L}{2\pi\xi_i(K,L)},
\end{eqnarray}
where $\xi_i (K,L)$ governs the decay of a correlation function $G_i(r)$.
According to the finite-size scaling\cite{fss}, 
the gap in the vicinity of a critical point scales as  
\begin{eqnarray}
X_{i}(K,L)&=&X_{i}+a_1(K-K_{\rm c})L^{y_t}+a_2(K-K_c)^2L^{2y_t}\nonumber\\
          &&+\cdots+buL^{y_u}+\cdots\label{Xh1fss},
\end{eqnarray}
where $X_{i}$ is the corresponding scaling dimension, which is related to an exponent $y_i=2-X_i$
according to the conformal invariance\cite{conformal}. $u$ is an irrelevant field, 
and $y_u<0$ is the corresponding irrelevant exponent.
$a_1$,$a_2$, and $b$ are unknown constants. 

We focus on the the energy-energy correlation $G_t(r) $ and two magnetic correlations
$G_A(r)$ and $G_B(r)$.
Generally speaking, the magnetic correlation function is defined as $G(r)=\langle s_0s_r\rangle$. 
For the HM model, since the spontaneous magnetizations on the two sublattices behave differently, we look two types of 
magnetic correlation functions: 
$G_A(r)$ with $s_0$ and $s_r$ on the A sublattice and 
$G_B(r)$ on the B sublattice. 

Let $\Lambda_1$ be the largest eigenvalue in the subspace that breaks the spin up-down symmetry, which means
that the associated eigenvector $\vec{v_1}$ satisfies
\begin{equation} 
\vec{v_1}=-{\mathbf F} \vec{v_1},
\end{equation} 
where ${\bf F}$ is the operator flipping spins. Thus, the scaled gap $X_1(K, L)$ is identified as $X_{h1}(K, L)$ and the
corresponding correlation is $G_A(r)$.

Let $\Lambda_2$ and $\Lambda_3$ be the second and the third largest eigenvalue in the subspace keeping the spin up-down 
symmetry. It is not {\it a priori} clear which of the two corresponding gaps is the thermal one $X_t(K, L)$ or the magnetic 
one $X_{h2}(K, L)$.
Based on the magnitudes of $X_t=2-y_t$ and $X_{h2}=2-y_{h2}$, we identify $X_2(K, L)$ and 
$X_3(K, L)$ as $X_{h2}(K, L)$ and $X_{t}(K, L)$, respectively.

\begin{figure}[htbp]
\includegraphics[scale=0.9]{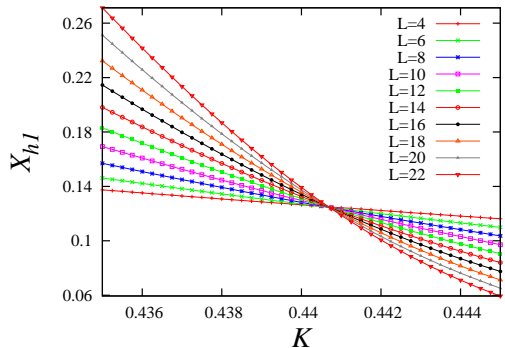}
\caption{(Color online) Scaled gap $X_{h1}(K,L)$ versus $K$ for a sequence of system size $L$ for the uniform HM model,
whose critical point is $K_{\rm c}=\log(1+\sqrt{2})/2=0.440687$.}
\label{xh1}
\end{figure}

\begin{figure}[htbp]
\includegraphics[scale=0.9]{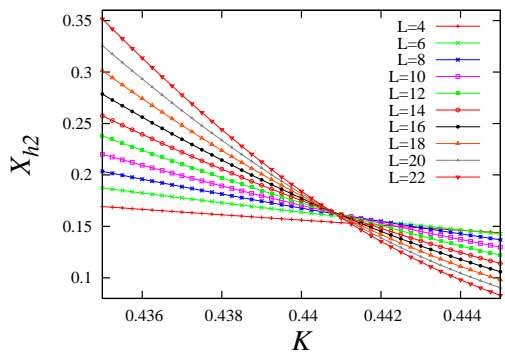}
\caption{(Color online) Scaled gap $X_{h2}(K,L)$ versus $K$ for a sequence of system size $L$ for the uniform HM model ($r=1$),
whose critical point is $K_{\rm c}=\log(1+\sqrt{2})/2=0.440687$.}
\label{xh2}
\end{figure}

Figures \ref{xh1} and \ref{xh2} illustrate $X_{h1}(K,L)$ and $X_{h2}(K,L)$ versus $K$ 
in the uniform case, respectively.

We then numerically solve the finite-size scaling equation
\begin{eqnarray}
X_{i}(K,L)&=&X_{i}(K,L-2), \label{xh1f}
\end{eqnarray}
for $i=h1, h2$, respectively.
The solution $K_{\rm c}(L)$ satisfies 
\begin{eqnarray} 
K_{\rm c}(L)=K_{\rm c}+a u L^{y_u-y_t}+\cdots.\label{Kcfss}
\end{eqnarray}
Here $a$ is an unknown constant. Since $y_u<0$ and $y_t\ge0$, $K_{\rm c}(L)$ converges to the critical point $K_{\rm c}$ with 
increasing system sizes.  (\ref{Kcfss}) is used to determine the critical point in our numerical procedure. 
For the uniform case, we obtain $K_{\rm c}=0.44068679(5)$, which is in good agreement with the exact 
solution $K_{\rm c}=\log(1+\sqrt{2})/2$. 
We have also estimated the critical points of the cases with $r=K^\prime/K=2, 3, 4$, and 5, where $K=K_1=K_3$ 
and $K^\prime=K_2=K_4$. 
Both  $X_{h1}(K, L)$ and $X_{h2}(K,L)$ have been used to estimate the critical points, 
we list the best estimations in Table \ref{ujtab}. 
Our numerical estimations of $K_{\rm c}$ 
are consistent with the theoretical results in a high accuracy. 

Exactly at the critical point $K_{\rm c}$, (\ref{Xh1fss}) reduces to 
\begin{eqnarray}
X_{i}(K_{\rm c}, L)&=&X_{i}+buL^{y_u}+\cdots,\label{Xh1cri}
\end{eqnarray}
which is used to determine the scaling dimensions $X_{h1}$ and $X_{h2}$. 
For the uniform case $r=1$, it gives $X_{h1}=0.125000(1)$, and $X_{h2}=0.166666(1)$, which are consistent with the analytic 
results (Table \ref{ujtab}).
We applied the same procedure to $r=2, 3, 4, 5$. 
The numerical estimations and the associated analytic results are consistent, see  Table \ref{ujtab}.

\begin{figure}[htbp]
\includegraphics[scale=0.95]{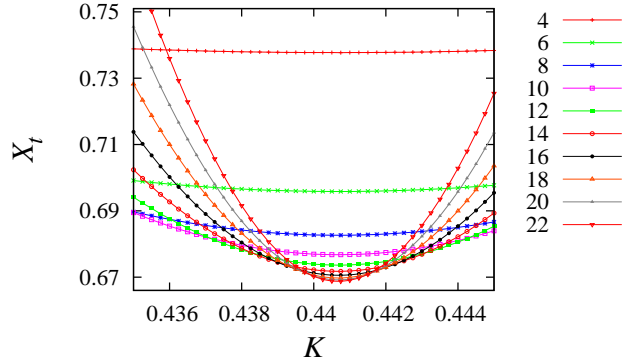}
\caption{(Color online) Scaled gap $X_{t}(K,L)$ versus $K$ for a sequence of system size $L$ for the uniform HM 
model ($r=1$), 
whose critical point is $K_{\rm c}=\log(1+\sqrt{2})/2=0.440687$.}
\label{xt}
\end{figure}

We also calculate the scaled gap $X_t(K,L)$ in the vicinity of critical point. The scaling behavior of $X_t(K, L)$ seems 
different from that of $X_{h1}$ or $X_{h2}$, as shown in Fig. \ref{xt} for the uniform case. This is the result of $a_1=0$
in Eq. (\ref{Xh1fss}). 
The location of the minimum of $X_t(K, L)$ for a given $L$ can be denoted 
as $K_{\rm c}(L)$, which converges to the critical point $K_{\rm c}$ when $L \to \infty$.  This property can also be used 
to estimate $K_{\rm c}$, e.g., as was done in \cite{henkcubic}. 
But we don't do this in current paper.
By calculating $X_t(K, L)$ at the estimated $K_{\rm c}$, we
obtain the scaling dimension $X_t$ according to (\ref{Xh1cri}).  
The results are listed in Table \ref{ujtab}.

The finite-size-scaling behavior of the free energy density
at the critical point determines the central charge $c$ 
according to \cite{fren1,fren2}
\begin{eqnarray}
f(L)\simeq f(\infty)+\frac{\pi c}{6L^2}. \label{ffss}
\end{eqnarray}
Fitting the data of the free energy density according to (\ref{ffss}), we obtain the central charge $c=1.000000(1)$ for 
the uniform case.
For the other cases, the numerical estimations of the central charge are listed in Table \ref{ujtab}.  
The central charge for all $r$ takes the fixed value $c=1$ as expected for transitions breaking four-fold symmetry. 

Furthermore, we consider the 4-state Potts point $r=r^*=3.3482581805\cdots$. 
A finite-size scaling analysis based on transfer matrix calculations
is performed at this point. The  numerical estimations obtained are listed in Table \ref{ujtab}. 
No logarithmic corrections are found, as in the Baxter-Wu model. 

\begin{ruledtabular}
\begin{table*}[hbtp]
\caption{Critical points, critical exponents, and scaling dimensions of the HM model. 
$r=K^\prime/ K$,  where $K=K_1=K_3$ and $K^\prime=K_2=K_4$. $r^*=3.3482581805\cdots$. T=Theoretical predictions, 
MC=Numerical result based on Monte Carlo simulations. TM=Numerical result based on transfer matrix calculations. The critical exponents 
and the scaling dimensions are related by $y_t=2-X_t$, $y_{h1}=2-X_{h1}$, and $y_{h2}=2-X_{h2}$.
}

 \begin{tabular}{r r|l|l|l|l|l|l}
      $r$  &     &     1        &      2      &    3        &   $r^*$     &     4      &    5        \\
\hline
     $K_{\rm c}$ & T   & 0.4406867935 & 0.3046889317&0.2406059125 &0.225147108  &0.2017629641&0.1751991102 \\
           & TM  & 0.44068679(5)& 0.3046889(1)&0.2406059(1) &0.2251472(2) &0.2017629(1)&0.1751991(2) \\
\hline
     $y_t$ & T   & 4/3          &1.39668184   &1.47604048   &3/2          &1.53960311  &1.58921160   \\
           & MC  & 1.332(2)     &1.397(2)     &1.478(3)     &1.500(1)     &1.540(3)    &1.590(2)    \\
     $X_t$ & T   & 2/3          &0.60331816   &0.52395952   &1/2          &0.46039689  &0.41078840  \\
           & TM  & 0.66666(1)   &0.603318(1)  &0.52396(1)   &0.50000(1)   &0.46040(2)  &0.41079(1)  \\
\hline
  $y_{h1}$ & T   & 15/8         &15/8         &15/8         &15/8         &15/8        &15/8         \\
           & MC  & 1.874(1)     &1.874(2)     &1.876(2)     &1.874(2)     &1.877(3)    &1.874(2)     \\
  $X_{h1}$ & T   & 1/8          &1/8          &1/8          &1/8          &1/8         &1/8          \\
           & TM  & 0.125000(1)  &0.125000(1)  &0.125000(1)  &0.12499(1)   &0.12500(1) &0.12498(3)    \\
\hline
  $y_{h2}$ & T   & 11/6         &1.84917046   &1.86901012   &15/8         &1.88490078  &1.89730290   \\
           & MC  & 1.833(1)     &1.848(2)     &1.870(2)     &1.875(1)     &1.886(2)    &1.896(4)     \\
  $X_{h2}$ & T   & 1/6          &0.15082954   &0.13098988   &1/8          &0.11509922  &0.10269710   \\
           & TM  & 0.166666(1)  &0.150829(1)  &0.130990(1)  &0.12500(1)   &0.115099(1) &0.102696(2)  \\
\hline
  $c$      & T   &1             & 1            &  1         &1            & 1          & 1           \\
           & MC  &1.000000(1)   & 1.000000(1)  &0.99998(3)  &1.0000(1)   & 1.0000(2)  & 0.9999(1)   \\
\end{tabular}
\label{ujtab}
\end{table*}
\end{ruledtabular}

\section{Monte Carlo simulations}
\label{sec_MC}
The Baxter-Wu model has been simulated by the Metropolis algorithm\cite{bwMC}, the Wang Landau algorithm\cite{wanglandau},
and a cluster algorithm\cite{henkbw,dengbw}.
The cluster algorithm is similar to the Swendsen-Wang algorithm\cite{sw} for the Potts model. 
In this algorithm, the triangular lattice is divided as three triangular sublattices. 
By randomly freezing the spins on one of the sublattices, the other 
two sublattices compose a honeycomb lattice with pair interactions, then a Swendsen-Wang type algorithm can be formed to update the spins on this honeycomb 
lattice. 
In current paper, we suitably  modify this cluster algorithm to simulate the HM model. 

\begin{figure}[htbp]
\includegraphics[scale=0.8]{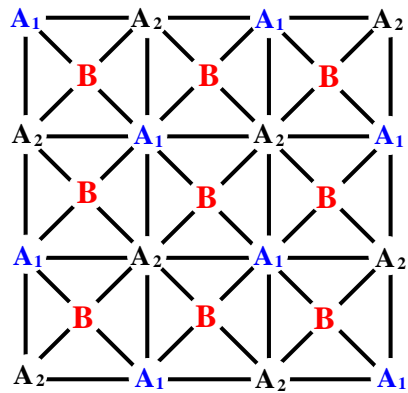}
\caption{ (Color online)
When ${\rm B}$ spins are frozen, the interactions of ${\rm A}$ spins (${\rm A_1}$ and ${\rm A_2}$ spins) reduce to pair interactions; 
when ${\rm A_1}$ spins are frozen, the interactions of ${\rm B+A_2}$ spins reduce to pair interactions; 
when ${\rm A_2}$ spins are frozen, the interactions of ${\rm B+A_1}$ spins reduce to pair interactions. 
}
\label{updatespin}
\end{figure}

The algorithm is divided into three steps:
\begin{enumerate}
\item Step 1, update the spins on sublattice A.

       Let the spins on sublattice B be unchanged (frozen), the interactions of the spins 
reduce to pair interactions on sublattice A.

       {\it A. bonds.} An vertical (horizontal) edge on sublattice A belongs to two triangles,  
the interactions of the left (upper) and right (down) triangles can be denoted $K_l$ and $K_r$ respectively, which may take 
the values $K_1, K_2, K_3$, or $K_4$.
Place a bond on this edge with probability 
$p=1-e^{-2K_l-2K_r}$ if both the products of spins in the left (upper) triangle and in the right (down) triangle are 1, 
$p=1-e^{-2K_l}$ if only the product in the left (upper) triangle is 1, 
$p=1-e^{-2K_r}$ if only the product in the right (down) triangle is 1, 
$p=0$ otherwise.

      {\it B. clusters.} A cluster is defined as a group of sites on sublattice A connected through the bonds. 

      {\it C. update spins.} Independently flip the spins of each cluster with probability 1/2.

\item Step 2, update the spins on sublattice B and half of the spins on sublattice A.
      
      This step is very similar to step 1, but we freeze only half of the spins on sublattice A, which are labeled ${\rm A_1}$ (Fig. \ref{updatespin}).
      The other spins (on sublattice B+${\rm A_2}$) are updated, whose interactions also reduce to pair interactions when ${\rm A_1}$ 
      spins are frozen.
\item Step 3, update the spins on sublattice B and the other half of the spins on sublattice A.

      In this step, the spins labeled ${\rm A_2}$ (in Fig. \ref{updatespin}) are frozen, other spins (on sublattice B+${\rm A_1}$) are updated.
\end{enumerate}
In a complete sweep, all the spins are updated twice.

In the simulations, the sampled variables include the specific heat $C$,  the magnetizations $M_{\rm A}$
on sublattice A, and $M_{\rm B}$ on sublattice B. 

The specific heat is calculated from the fluctuation of energy density $E$
\begin{eqnarray}
C=\frac{L^2(\langle E^2 \rangle-\langle E \rangle ^2)}{k_BT^2},
\end{eqnarray}
where $L$ is the linear size of the system, and $\langle\cdots\rangle$ means the ensemble average.
The magnetizations are defined as 
\begin{eqnarray}
M_{\rm A}&=&\frac{\big\langle\big|\sum\limits_{i=1}^N \sigma_i \big|\big\rangle}{N},\\
M_{\rm B}&=&\frac{\big\langle\big|\sum\limits_{i=1}^N s_i\big| \big\rangle}{N},
\end{eqnarray}
where $N=L^2$ is the number of total sites on sublattice A or B. 

In order to demonstrate our numerical procedure based on the Monte Carlo simulations, we take the uniform case ($K=K_1=K_2=K_3=K_4$) 
as an example. 
The cluster algorithm is very efficient, which easily allows us to do meaningful simulations for system with linear size up to $L=384$.
All the simulations are performed at the critical point $K_{\rm c}$. After equilibrating the system, $10^7$ samples were taken for each 
system size. 
Using the Levenberg-Marquardt algorithm, we fit the data of $C$ according to the finite-size scaling formula\cite{fss}
\begin{eqnarray}
C(L)=C_0+L^{2y_t-d}(a+bL^{y_1}),
\end{eqnarray}
where $bL^{y_1}$ are the leading correction-to-scaling term with $y_1<0$ the leading irrelevant exponent. $d=2$ is the 
dimensionality of the lattice. 
$C_0$, $a$, and $b$ are unknown parameters.
The fitting yields $y_t=1.332(2)$, which is in good agreement with the exact result $y_t=4/3$.

\begin{figure}[htbp]
\includegraphics[scale=1.0]{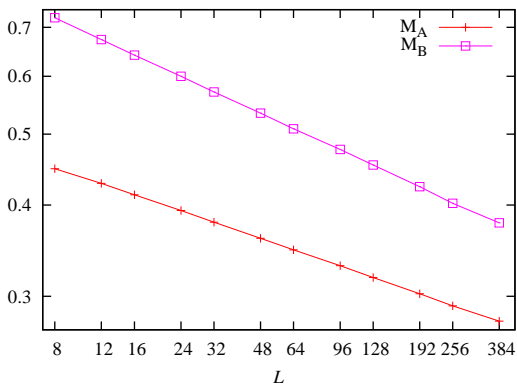}
\caption{(Color online) Log-Log plot of the magnetizations $M_{\rm A}$ and $M_{\rm B}$ versus system size 
$L$ for the HM model at the critical point $K_{\rm c}=\log(1+\sqrt{2})/2$ for the uniform case ($K=K_1=K_2=K_3=K_4$).}
\label{mags}
\end{figure}

The finite-size scaling behaviors of the magnetizations $M_{\rm A}$ and $M_{\rm B}$ are
\begin{eqnarray}
M_{\rm A}&=&L^{y_{h1}-d}(a+bL^{y_1}),\label{fsma}\\
M_{\rm B}&=&L^{y_{h2}-d}(a^\prime+b^\prime L^{y^\prime_1}),\label{fsmb} 
\end{eqnarray}
respectively.
The log-log plot of $M_{\rm A}$ and $M_{\rm B}$ versus $L$ are shown in Fig. \ref{mags}.
The fits yield $y_{h1}=1.874(1)$, and $y_{h2}=1.833(1)$, where $y_{h1}$ and $y_{h2}$ are 
in good agreement with the analytic predictions $y_{h1}=15/8$ and $y_{h2}=11/6$, respectively.

We also simulate the $r=2, 3, 4, 5$, and $r^*$ cases.
The numerical estimations of $y_t, y_{h1}, y_{h2}$ and the corresponding analytic predictions are listed in 
Table \ref{ujtab}. They are in good agreement. 
In these fits, no logarithmic corrections to scaling are found. 

\section{Fractal structure of the model}
\label{sec_FC}
In addition to calculate the critical exponents $y_t$, $y_{h1}$, and $y_{h2}$, we also 
investigate the geometry properties of the model. 

The configurations of the HM model are represented by Ising spins. Thus we can define
``Ising clusters" for the model as in the Ising model\cite{IsingCluster}. 
To do so,  each sublattice is viewed as a square lattice.
For two nearest-neighboring Ising spins on a sublattice, they are considered to be in the same cluster if they have 
the same sign. Here, in saying ``nearest-neighboring",  the neighborhood of the site 
is the same to that of a site in a square lattice.
This definition is applied to the two sublattices respectively, thus there are two
types of Ising clusters. For clarity, we call the Ising cluster on sublattice A as ``A cluster", while the Ising cluster 
on sublattice B as ``B cluster". 

\begin{figure}[htpb]
\includegraphics[scale=1.0]{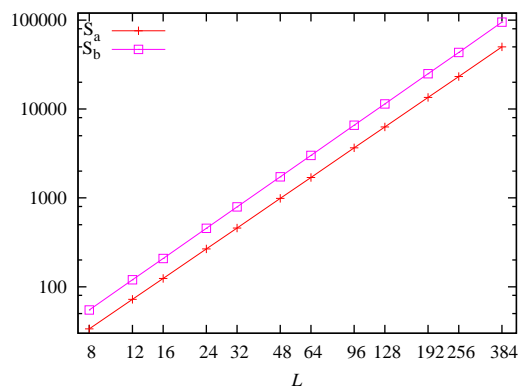}
\caption{(Color online) Log-Log plot of $S_a$ and $S_b$ versus system size $L$ for the HM model at the critical
point $K_{\rm c}=\log(1+\sqrt{2})/2$ for the uniform case ($K=K_1=K_2=K_3=K_4$).}
\label{smax}
\end{figure}

The simulations are also performed at the critical points. The size of the largest
A cluster and the largest B cluster are denoted $S_a$ and $S_b$, respectively. They satisfy the finite-size scaling 
\begin{eqnarray}
S_a&=&L^{D_a}(a+bL^{y_1}),\label{sa}\\
S_b&=&L^{D_b}(a^\prime+b^\prime L^{y^\prime_1}), \label{sb}
\end{eqnarray}
which means that the largest A cluster and the largest B cluster are fractals,
with $D_a$ and $D_b$ the fractal dimensions, respectively.
Figure \ref{smax} is an illustration of $S_a$ and $S_b$ versus system size $L$ at the uniform case.
Fitting the data according to (\ref{sa}) and (\ref{sb}), we find $D_a=1.888(1)$ and $D_b=1.925(1)$.  

The same procedure were applied to $r=2, 3, 4, 5$ and $r^*$. The results obtained are listed in 
Table. \ref{ujfrc}.  
The fractal dimension of the A cluster has a fixed value $D_a=1.888(2)$, 
while that of the B cluster varies with the ratio $r$. 

\begin{ruledtabular}
\begin{table*}[hbtp]
\caption{
Fractal dimensions of the HM model. $D_a$ is the fractal dimension of A cluster, $D_b$ is the fractal dimension of B cluster.
$r=K^\prime/ K$,  where $K=K_1=K_3$ and $K^\prime=K_2=K_4$, $r^*=3.3482581805$.}
 \begin{tabular}{r|c|c|c|c|c|c}
      $r$  & 1  & 2  & 3  & $r^*$  & 4 &5\\
\hline
     $D_a$ & 1.888(1) &1.889(1)&1.890(3)&1.889(2) &1.888(1) &1.888(2)\\
\hline
     $D_b$ & 1.925(1) &1.931(1)&1.939(1)&1.941(1) &1.945(2) &1.951(2) \\
\end{tabular}
\label{ujfrc}
\end{table*}
\end{ruledtabular}

\section{Conclusion and discussions}
\label{sec_Con}
In conclusion, we have numerically studied the critical properties of the HM model by means of finite-size scaling 
analysis based on the transfer matrix calculations and the Monte Carlo simulations. For the critical points and the 
critical exponents $y_t$ ($X_t$), $y_{h1}$ ($X_{h1}$),
 and $ y_{h2}$ ($X_{h2}$), our numerical estimations are in good agreement with the corresponding 
analytic predictions\cite{hintermann,wubw}. The analytic prediction for $y_{h2}$ is based on the conjectured spontaneous
polarization of the eight-vertex model, which remains unproved. Our numerical results verified the correctness of the prediction. 

In addition the central charge of the model is found to be $c=1$.
This is consistent with the fact that $y_{h1}=15/8$, 
$y_t$ and $y_{h2}$ vary continuously with the parameters, which is
related to the four-fold degeneracy of the ground states of the model. 
Usually, logarithmic corrections\cite{log,lvAT,j1j2-guo} show up in the 4-state Potts point for the model with 4-fold degenerate
ground states due to the second temperature field. However, 
we do not see such logarithmic corrections in the 4-state Potts point of the HM model. This is the same as the Baxter–Wu model, in which 
the amplitude of the marginally irrelevant operator is zero.

Furthermore, two unpredicted critical exponents are found to describing geometry properties of the model.  
We define clusters based on the Ising-spin configurations on the two sublattices of the model. 
The fractal dimension of the largest cluster on sublattice A takes fixed value $D_a=1.888(2)$, while the fractal dimension of the largest 
cluster on sublattice B varies continuously with the parameter of the model. 

\section*{Acknowledgment}
Ding thanks F. Y. Wu  for valuable discussion in terms of understanding the analytical results of HM model.
This work is supported by the National Science Foundation of China (NSFC) under Grant No.~11205005 (Ding) and 11175018 (Guo).

\end{document}